\documentclass[3p,preprint,11pt]{elsarticle}

\usepackage[brazil,english]{babel}
\usepackage[utf8]{inputenc}
\usepackage{float,epsfig}
\usepackage{amssymb}
\usepackage{url}
\usepackage[normalem]{ulem}
\usepackage[usenames]{color}

\DeclareUrlCommand\path{\urlstyle{sf}}

\newcommand{\definicao}[1]{{\it #1}}
\newcommand{\etAl}{{\it et al.}}

\journal{Elsevier}

\begin{document}


\begin{frontmatter}
  \title{SCTE: An open-source Perl framework for testing equipment control and data acquisition}

  \author[york,IF]{Luiz C. Mostaço-Guidolin\corref{cor1}}
  \ead{lm.guidolin@gmail.com}
  \address[york]{%
    Department of Mathematics and Statistics -- York University, 
    4700 Keele St., Toronto, ON, Canada M3J 1P3.
  }
  \cortext[cor1]{Corresponding author}

  \author[utfp]{Rafael B. Frigori}
  \address[utfp]{%
    Universidade Tecnológica Federal do Paraná, %
    Rua Cristo Rei 19, CEP 85902-490 Toledo, PR, Brazil. %
  }

  \author[IF]{Leonid Ruchko}
  \author[IF]{Ricardo M. O. Galvão}
  \address[IF]{%
    Departamento de Física Aplicada, IF --  Universidade de São Paulo, %
    Rua do Matão, Travessa R, 187. CEP 05508-090 São Paulo, SP, Brazil.%
  }


\begin{abstract}
  SCTE intends to provide a simple, yet powerful, framework for building data acquisition and equipment control systems for experimental Physics, and correlated areas. Via its SCTE::Instrument module, RS-232, USB, and LAN buses are supported, and the intricacies of hardware communication are encapsulated underneath an object oriented abstraction layer. Written in Perl, and using the SCPI protocol, enabled instruments can be easily programmed to perform a wide variety of tasks. While this work presents general aspects of the development of data acquisition systems using the SCTE framework, it is illustrated by particular applications designed for the calibration of several in-house developed devices for power measurement in the tokamak TCABR Alfvén Waves Excitement System.

\end{abstract}


\end{frontmatter}

\section*{PROGRAM SUMMARY}  

\begin{footnotesize}
  \begin{description}
    \item[{\em Authors:}] Luiz C. Mostaço-Guidolin 

    \item[{\em Program Title:}] SCTE 

    \item[{\em Journal Reference:}] 
    
    \item[{\em Catalogue identifier:}] 
    
    \item[{\em Licensing provisions:}] GNU/GPL version 3 

    \item[{\em Programming language:}] Perl version 5.10.0 or higher
    
    \item[{\em Computer:}] PC
    
    \item[{\em Operating system:}] GNU/Linux (2.6.28-11), should also work on any Unix-based operational system.
    
    \item[{\em Supplementary material:}] SCPI capable digital oscilloscope, with RS-232, USB, or LAN communication ports, null modem, USB, or Ethernet cables.
    
    \item[{\em Keywords:}] Perl, automation, data acquisition, equipment control,
          test instrument, TCABR, SCPI.

    \item[{\em PACS:}] 01.50.hv, 01.50.Lc, 01.50.Pa. 
    
    \item[{\em Classification:}] 4.14 
    
    \item[{\em External routines/libraries:}] Perl modules: {\tt Device::SerialPort}, {\tt Term::ANSIColor}, {\tt Math::GSL}, {\tt Net::HTTP}
    
    \item[{\em Subprograms used:}] Gnuplot 4.0 or higher
    
    \item[{\em Nature of the problem:}] Automation of experiments and data
    acquisition often requires expensive equipment and in-house development of
    software applications. Nowadays personal computers and test equipment come
    with fast and easy-to-use communication ports. Instrument vendors
    often supply application programs capable of controlling such devices, but
    are very restricted in terms of functionalities. For instance, they are not
    capable of controlling more than one test equipment at a same time or to
    automate repetitive tasks. SCTE provides a way of using auxiliary equipment
    in order to automate experiment procedures at low cost using only free, and 
    open-source operational system and libraries.

    \item[{\em Solution method:}] SCTE provides a Perl module that implements
    RS-232, USB, and LAN communication allowing the use of SCPI capable instruments
    [1]. Therefore providing a straightforward way of creating automation and data
    acquisition applications using personal computers and testing
    instruments[2].

    \item[{\em Program Summary References:}] \ \ 
        \begin{description}
            \item[] [1] SCPI Consortium, \definicao{Standard Commands for
            Programmable Instruments}, 1999. \\
             \url{http://www.scpiconsortium.org}.

            \item[] [2] L. C. B. Mostaço-Guidolin, Determinação da configuração de ondas de Alfvén excitadas 
              no tokamak TCABR, Master's thesis, Universidade de São Paulo
              (2007).\\
              \url{http://www.teses.usp.br/teses/disponiveis/43/43134/tde-23042009-230419/}
         \end{description}
  \end{description}
\end{footnotesize}

\section{Introduction}  

Data acquisition systems tend to be specialized, task-specific, and
mission-critic components in laboratories of experimental Physics, requiring
a great deal of coordination to develop, document and maintain.  They are the
command centre of the most diverse environments composed of cables, computers,
custom electronics, specialized instruments, detectors, communication protocols,
and interface buses.  Building such a critical system is always a challenging
problem.  The design of a data acquisition system must take into account the
characteristics of the individual experiment.  Despite the uniqueness of each
system, common requirements can be factorized out.
According to Gutleber \etAl~\cite{Gutleber2003}, such factors are divided into functional and non-functional requirement domains, which establishes desired characteristics of a data acquisition system.  The functional requirement domain may be further divided into six subcategories in terms of (i) communication, (ii) device access, (iii) configuration, (iv) control and monitoring, (v) re-usability of application modules, and (vi) user accessibility requirements.  In the non-functional domain, four other subcategories are identified, namely maintainability and portability, scalability, flexibility, and component identification (see reference \cite{Gutleber2003} for more details).

While data acquisition systems are often developed for large and complex
laboratory environments and designed to control state-of-the-art experimental
setups, other less critic, smaller, and repetitive task could benefit from a
small-scaled data acquisition system capable of controlling testing instruments,
performing autonomous calibrations or even stress testing custom electronic devices
prior to their incorporation into the main system.  Since the creation of the
\definicao{Standard Commands for Programmable Instruments} language
(SCPI)\cite{SCPIConsortium1999,Geathers}, which defines a common interface
language between computers and test instruments, the instrumentation industry
has been massively adopting this language in their equipment.  As a result, it
became possible to develop applications capable of communicating and controlling
a vast gamut of instruments directly from a personal computer, and using a variety of
buses.

The SCPI language is comprised of hierarchically structured commands encoded
in ASCII text strings~\cite{SCPIConsortium1999}. These commands are sent to the
instruments in order to either perform a wide variety of set or query operations, like instructing a device to perform a self calibration, or reading the frequency of a signal.  Moreover, the SCPI standard was designed to be vendor,
interface, and instrument independent, thus it may be employed to control any
capable instrument.

The framework application presented in this work, was devised to take
advantage of the unexplored facilities created by the SCPI definition.  The
\definicao{Software for Controlling Testing Equipment} (SCTE) is a framework developed in the Perl programming language \cite{Perl,PerlSite}. It was created in 2004 as an open
source project, designed to automate the process of calibration of a set of
devices for determining the radio-frequency power applied to the plasma by the
\definicao{Tokamak Chauffage Alfvén Brésilien's - Alfvén Waves Excitement
System} (TCABR-AWES)~\cite{RuchkoAASAWH,MsMostacoGuidolin}, but it was later refactored and remodelled as
a framework for building general data acquisition and instrument control
systems.

Initially designed to use only the RS-232 (serial)
bus~\cite{serial}, it has been extended to support the \definicao{Universal Serial Bus} (USB) through the \definicao{USB Test \& Measurement Class} (USBTMC),
which is supported in GNU/Linux systems as the kernel module {\tt usbtmc.ko}~\cite{AgilentUSBTMC,AgilentUSBTMC2}, as well as Ethernet LAN support. Since the SCPI language is
bus independent SCTE may be adapted to use legacy \definicao{General Purpose
Interface Bus} (GPIB).
Developed under GNU/Linux systems, it has
been licensed as free and open-source software (FOSS), uses only FOSS
libraries, and satisfies most of the recommendations stated by Gutleber \etAl~\cite{Gutleber2003}.
The GNU/Linux
system, as any other Unix-like systems, is a stable platform that provides a
convenient layer for interfacing with devices through its {\tt /dev} directory,
which makes the communication with external devices easier~\cite{tanenbaum2008modern}.

Perl is an interpreted programming language with object
orientation capabilities, portable, embeddable, has low defect density,
is considered free of known security defects, has full integration with most database
management systems, and is very powerful for text manipulation, which makes it ideal for working with SCPI commands \cite{PerlSite}. The use of interpreted languages avoids handling memory allocation and dealing with complex data structures, which ultimately accelerate the development process when compared to the C language at a small expense of performance loss. Additionally, it is a well stablished language, and is vastly present in the TCABR data acquisition system \cite{TCAqs}. 

SCTE has been tested with Tektronix oscilloscopes (series TDS-200, TDS-1000,
TDS-2000, and TDS-3000)~\cite{TektronixTDS}, and with Agilent oscilloscopes (model DSO1012A)~\cite{AgilentDSO}, using RS-232, USB, and Ethernet ports. Considering that one of the aims of the SCPI standard is interoperability, any SCPI capable device is prone to be used with the SCTE framework.

\subsection{Data Acquisition and Control Systems Overview: SCTE framework}  

Data acquisition and control systems are commonly developed for large and
complex projects, including particle
accelerators and colliders, cosmic ray detectors arrays, and tokamaks~\cite{TCAqs,LHC,Auger,Valcarcel2009,Goncalves2010}.  In these setups the number of devices under
control, data acquisition rates, signals frequencies, amount of data storage
space, computational power, and even geographical distances may be overwhelming
demanding the coordination of the efforts of a huge number of people.  Even in
small-scale environments the requirements may be quite challenging.  A common
characteristics to many, if not all, experimental physics setups is the use of
measurement instruments, for providing input signal, controlling
the experiment, and making measurements. Frequently, such experiments require in-house developed devices, which will require to be calibrated and tested, a task that may be time consuming, and thus ideal for automation. SCTE has been designed to be a flexible framework that enables fast development of data acquisition and control systems.  It can be applied to small-scale setups, or be easily integrated as an auxiliary data acquisition system for larger environments.
 
Most modern commercial instruments are equipped with communication ports
and are SCPI capable.  Commercial software packages, LabView and Matlab for instance, may be employed in the creation of simple data acquisition systems, however they can cost a substantial amount of money for a single license, and demand much more powerful computer machines to run than SCTE. Therefore, can be easily replaced by SCTE.

\begin{figure}[H]
	\begin{center}
		\includegraphics[width=0.75\columnwidth,angle=0]%
            {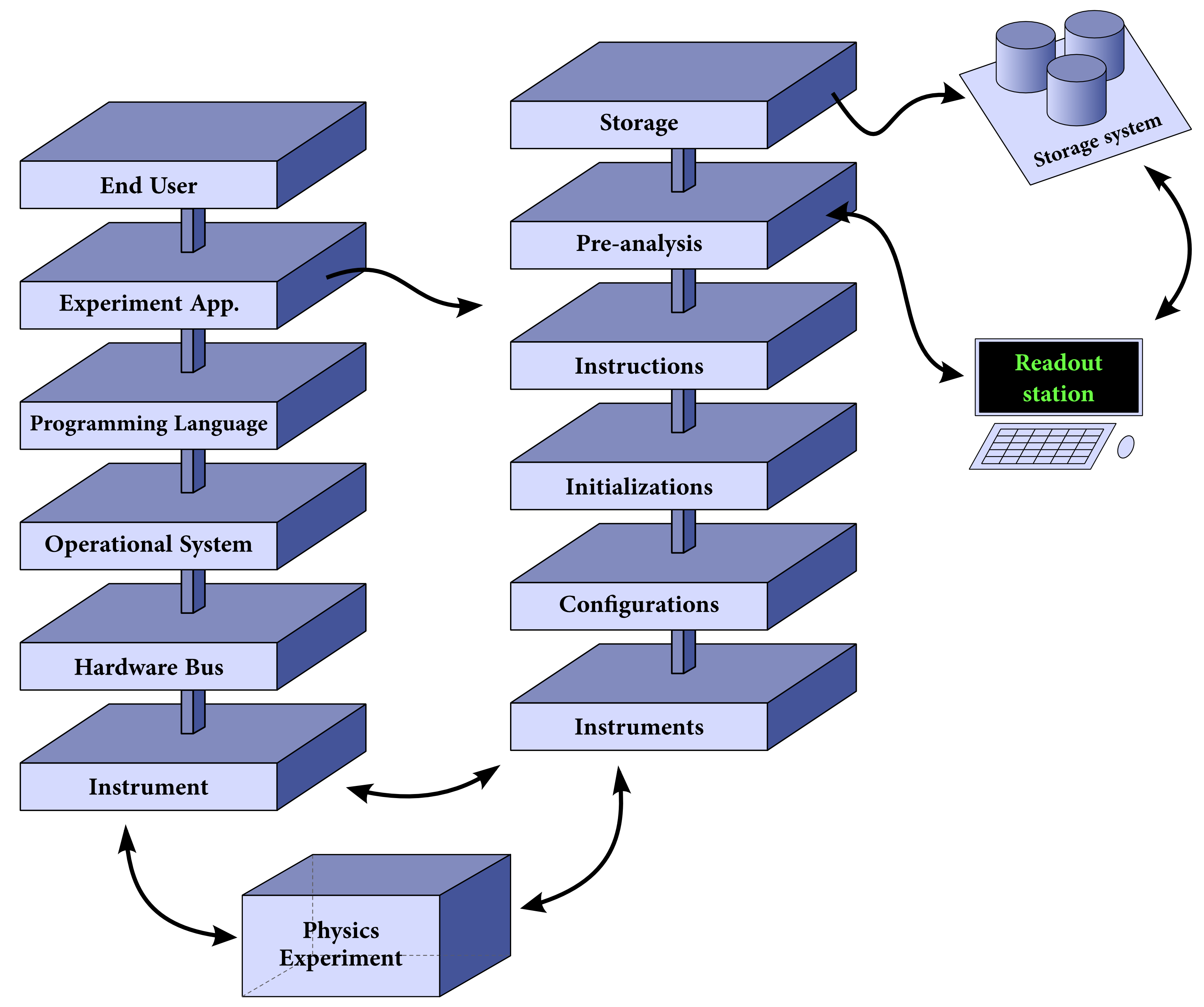}
  		\caption{Layered model of a general data acquisition and control %
  		  system, based on the SCTE framework.}
  		\label{fig:layered_model}
	\end{center}
\end{figure}

The general problem of developing a data acquisition and control systems for
experiments in Physics can be modelled as shown in
Figure~\ref{fig:layered_model}.  In this model there are two sets of layers, one
describing the technical aspects of the system, including the instruments involved
in the experiment, the way they are connected to the controlling system, and the
programming language to be adopted.  The second set of layers describes the
logical aspects of the development of the experiment application, including the manipulation of the instruments, configurations, and storage management.

 \vspace{-0.25cm}

Thinking a data acquisition system begins with the identification of the
necessary instruments required to measure or quantify all the phenomena involved in
the experiment. Considering the leftmost stack of layers in
Figure~\ref{fig:layered_model}, once the required instruments are identified,
other components become automatically determined i.e., hardware bus and
communication protocol. Hardware bus may be RS-232, USB, GPIB, or LAN.
With the bus identified, one or two communication protocols may be necessary, a
transport protocol, and a language protocol.  The transport protocol is
responsible for transmitting the language protocol messages over the media. In
the case of a device connected via LAN, transport and routing protocols, like
TCP/IP, may be employed to deliver the language protocol messages. In this cases, it is also common the use of application protocols such as HTTP to transmit the language protocol messages~\cite{Zimmermann1980,OSI_ISO}. In cases of devices connected via USB, for example, no transport or routing protocols are necessary and the instrument access is directly facilitated by the operating system and the language protocol messages (SCPI commands) can be sent directly to the device.

In terms of operating systems (OS), stability, security, openness, efficiency,
and device access, are the most important features that must be taken into account.
Data acquisition systems are mission critic, frequently running for long periods
of time, and controlling sensible and expensive devices. Therefore, the
operating system it relies on should be able to run for long periods without
system crashes, interruptions, or instabilities that may result in the
interruption of service, or even damage to the equipment. For these reasons,
security is equally important as stability since security failures, like
computer viruses, may jeopardize the entire environment. Furthermore, it is
desirable that data acquisition systems, and any of its components, must remain completely isolated from the internet in its own private network, dedicated solely to the data acquisition system, thus avoiding "hacker" attacks~\cite{LHCattack}.

Operating system openness relates to security in the sense that security
flaws may be rapidly identified and addressed, but also to the ability of
adapting the OS to the specific needs, as building an embedded kernel specific for devices that are part of an array of detectors, providing only the
necessary functionalities. In terms of efficiency, it is desirable to
choose an OS with the lowest overhead as possible, and which does not require
graphical user interfaces to run, thus reducing hardware requirements, increasing stability, and improving security~\cite{GUIsecurity}.

The next layer in the stack corresponds to the programming language. At this
point there is no consensus on what is the most appropriated language, in view of the fact that many factors have to be considered and only one language may not be appropriated for the entire environment. Although, well established languages are recommended over newer and trendy ones.

The experiment application is the end result that combines all the components of
a data acquisition system and makes it useful to the end user. It is comprised of
the instruments and all necessary manipulations that makes it useful. Every data
acquisition system has to cope with all the aspects presented in the rightmost
stack in Figure~\ref{fig:layered_model}, which not only presents all necessary
components of such application, but also the logical structure of a data
acquisition system.

Instruments are the central part of a data acquisition system. Therefore, it is desirable to have a framework to handle the intricacies of communicating with the devices, in order to manipulate any number of
instruments with the lowest complexity possible. At this stage, the
support for handling such instruments must be added. Using object oriented
languages favours the control of several instruments at a low complexity cost,
since each instrument may be treated as an instance of a class of instruments.
Following the addition of the instruments to the application is the
configuration step, which involves reading configuration files for the
experiment configuration itself, and reading the configuration parameters
for the instruments. The actual configuration of the instruments is performed in
the initialization part of the experiment. At this stage, the instruments must
be initialized, which involves testing the communication channels and sending
the configuration parameters. Still in the initializations layer, variables,
file handlers, database connections should be initialized as well. After this
pre-experiment phase, the instructions phase comes, which corresponds to the
experiment itself i.e., the sequence of operations to control de instruments,
fire systems, perform readings, and acquiring data. During the experiment phase,
or after it, depending on the necessity, pre-analysis routines and storage
procedures may be carried out. The pre-analysis may be employed to provide
monitoring and controlling capacity to the application, preferably using readout
stations. It is desirable that the storage of the information be handled by
dedicated systems and database management systems, completing the setup environment
for a data acquisition.

When dealing with multiple instruments, or extending SCTE's applicability to larger 
environments with more complex workflows, the requirements of the application increases. Although SCTE is an interface to instruments, it does not account for aspects as concurrence, independent timing sequences, interlocks between systems, triggering, event handling, network latency, among other elements. These aspects, therefore, are left on the control of the developer of the application.

\section{Software Overview} 

SCTE has been developed following the premisses of the model presented in
Figure~\ref{fig:layered_model}. Therefore, it is divided in two parts, a 
module for communicating with instruments, and a group of examples of data
acquisition systems. The communication module fits in between the Experiment 
App. and Programming Language layers, at the leftmost stack in 
Figure~\ref{fig:layered_model}. The systems provided as examples, present 
implementations adopting the structure on the right-most stack of the same
figure.

\subsection{Communication Module}  

\begin{figure}[H]
	\begin{center}
		\includegraphics[width=0.75\columnwidth,angle=0]%
            {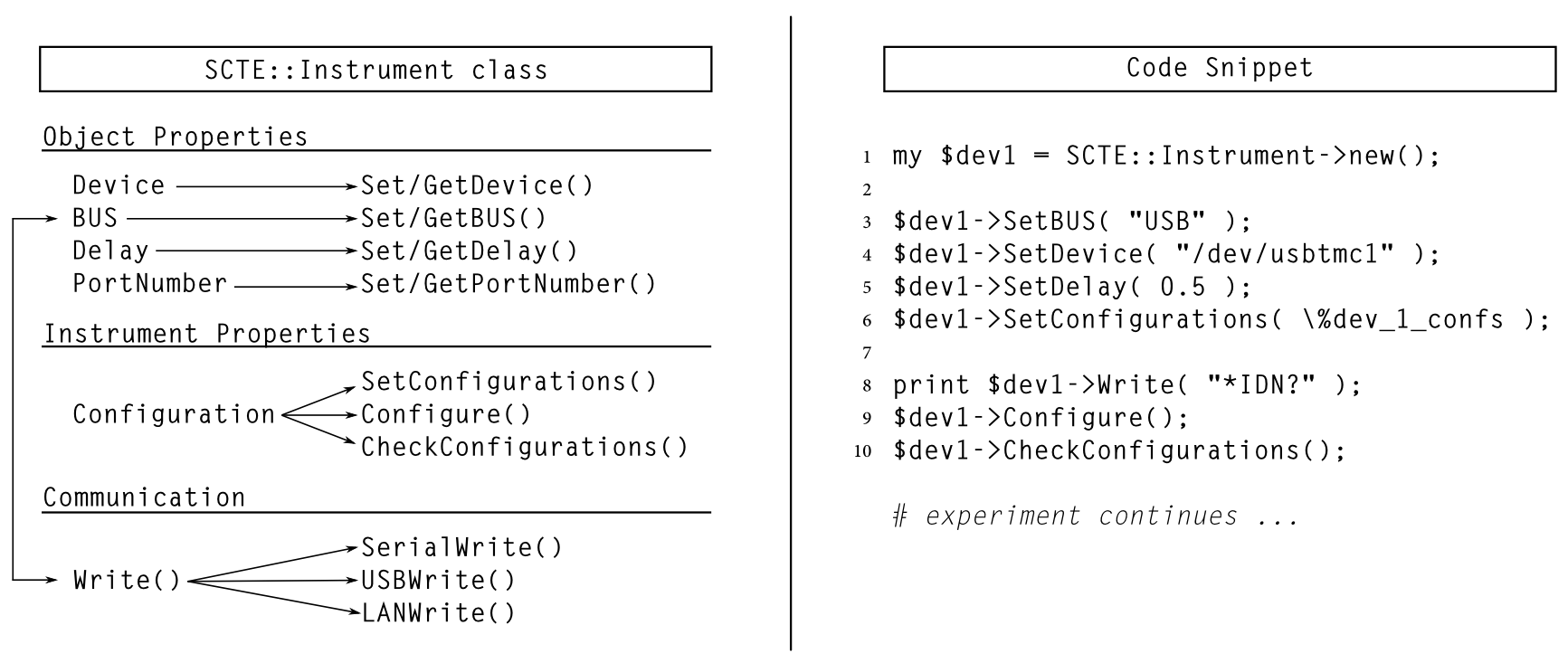}
  		\caption{{\tt SCTE::Instrument} class and usage example.}
  		\label{fig:scte_instrument}
	\end{center}
\end{figure}

The communication module, named as {\tt SCTE::Instrument}, is a perl module that
implements a class with the same name. Utilizing the object oriented programming
paradigm, this class encapsulates the intricacies associated to the
communication using different buses. With this approach, an object is associated to an instrument, has its own configuration parameters, but shares the
same communication interface established by the methods of the SCTE::Instrument class, despite
the particular bus that each instrument is connected to
(Figure~\ref{fig:scte_instrument}). The developer working in the
development of the data acquisition system is able to manipulate several
instruments at a time, each one via its own object, and abstract from the
problem of dealing with the communication interface for each instrument.
Therefore, focusing solely on the main issues, as developing the data acquisition system.

 \vspace{-0.25cm}

Each {\tt SCTE::Instrument} object requires four parameters: the device address (either the path to the device file to which the instrument is interfaced, or its IP address in the network), bus type, delay between sending an instruction and starting reading the response, and instrument specific configuration parameters. On supporting RS-232 (serial), USB, and Ethernet (LAN) buses, the developer must inform the instrument object: the bus type to which the device is connected to (via {\tt SetBUS()} method), thus allowing the {\tt Write()} method to use the proper communication bus.

The access to the hardware is handled by the Linux kernel via its {\tt udev} file
system, which provides a convenient way of accessing devices connected directly to 
the machine by associating files to devices in the {\tt /dev} directory.  The
serial port is, in general, associated to {\tt /dev/ttySN} files, in which
{\tt N} corresponds to the number associated to a specific serial port. The
manipulation of such devices is facilitated by the use of the perl module {\tt
Device::SerialPort}, available from the \definicao{Comprehensive Perl Archive
Network} (CPAN)~\cite{cpan,CPANdeviceSerialPort}. USB instruments are
accessed via device files such as {\tt /dev/usbtmcN}, in which {\tt N}
corresponds to the number of an specific USB port. The support to this
kind of device is made available by the {\tt usbtmc.ko} kernel module, which is
provided by default with the Linux kernel in most of the currently available
GNU/Linux distributions. Finally, Ethernet support is provided by the perl module {\tt Net::HTTP}, which uses the IP address of the instrument and sends the SCPI commands via HTTP GET method.

The configuration parameters for the device are passed to the {\tt
SCTE::Instrument} object using the {\tt SetConfigurations()} method, which receives
a hash with the appropriate SCPI command strings. Since an SCPI command may be
used to set or query a parameter by simply adding a question mark to the end of
the command, the configuration parameters hash has these SCPI commands (without
question marks) as keys, and the respective SCPI command values to be set, as
values. In this way, the {\tt Configure()} method is used to send these
configuration parameters to the instrument and the {\tt CheckConfigurations()}
method can be used to query the configurations set in the instrument, and compare
to the defaults specified in the configuration hash. After setting the
configuration parameters the experiment application developer can simply send
SCPI commands to the instrument using the {\tt Write()} method with the command
string as argument and reading its return, which corresponds to the return message
sent by the instrument. A code snippet on the use of this class is shown in Figure ~\ref{fig:scte_instrument}.

\subsection{Experiments Bundle}  

Designing an experiment is equally, or maybe even more complex than designing
the instrument communication system itself. It may involve the coordination of
several instruments, computers, storage resources, databases, data
analysis and processing routines, handling failures, designing the user
interface, among many other issues.

Data acquisition systems for large experiment
setups, often require graphical user interfaces (GUI), which enables the end
user to more easily manipulate the system. The GUI interface is employed
whenever the complexity of the setup needs to be reduced, or it is desirable to add
data analysis functionalities in it. For simpler experiments, like
calibrations or stress testing of a newly in-house developed device, a simple text-based user interface (TUI) may be much more effective and faster to develop. It also enables the user to manipulate the application remotely via a secure shell, or to build simple web-interfaces for monitoring purposes. In this way, the example provided with SCTE are built with TUIs and make external calls to a
plotting application for offering visualization of the data being acquired. The application of choice was Gnuplot, due to its simplicity and
broad adoption by the scientific community, although any other application for instance R or xmgrace, may be used as well.

SCTE package bundles four example experiments, together with the
{\tt SCTE::Instrument} module. These experiments aim to provide examples on how to develop data acquisition systems using SCTE. In these experiments, the basic functionality and usage of the SCTE communication layer and the integration with the SCPI language is presented. Experiments 1 and 2 are examples that use one and two devices respectively, in order to perform the same task. The purpose of Experiment 1 is to present the simplest case involving the use of only one instrument, while Experiment2 extends this design by using two instruments at the same time, thus exemplifying the manipulation of more than one device within the same application. Although SCTE has no known limitation regarding the number of instruments it can handle, it is limited by the number of devices supported by the USB and serial buses together.

The remaining examples (experiments 3 and 4) are simple demonstrations on how
to use SCTE to perform two specific tasks as acquire an snapshot of the
oscilloscope screen, and to acquire the wave form data points of the signal
being displayed by the oscilloscope. The aim of these examples is to present how SCTE can be employed to explore other functionalities of the instruments. The snapshot functionality may be used for documentation purposes, and capturing the wave form data points enables a more detailed analysis of the signal form. Although these functionalities were included in separate examples, they can be easily integrated to any data acquisition application at convenience. The experiments will be explained in further details in section~\ref{sec:exp_desc}.

\section{Installation}  

SCTE is comprised by two parts namely the communication module {\tt SCTE::Instrument}, which requires installation as any other perl module, and the experiments bundle. However, the experiments are stand alone applications that do not require installation, but depend on the {\tt SCTE::Instrument} and other third-party modules in order to properly run.
Detailed instructions concerning all the aspects of the installation and usage of SCTE, including the installation of other dependent modules, are provided in {\tt
README} files that accompany the SCTE package. Additionally, further instructions may be obtained in the SCTE project web site\footnote{\url{http://code.google.com/p/scte/}}, which contains links for
downloading SCTE, a wiki page that covers aspects of the installation, usage, and preparation of the operational system.

In order to be able to install the module {\tt SCTE::Instrument} and run the
experiment scripts, the perl modules {\tt Term::ANSIColor} , {\tt
Device::SerialPort}, {\tt Math::GSL}, and {\tt Net::HTTP} must be installed, as
 well as the plotting application
Gnuplot~\cite{cpan,CPANdeviceSerialPort,CPANtermAnsiColor,CPANGSL,CPANdeviceNetHTTP,gnuplot}.
Perl modules are available for download and installation via CPAN, although several GNU/Linux distributions provide packages to directly install these modules, this is also true for the Gnuplot application.

\section{Experiments Description}  
\label{sec:exp_desc}  

\begin{figure}[H]
  \centering
    \includegraphics[width=1\textwidth]{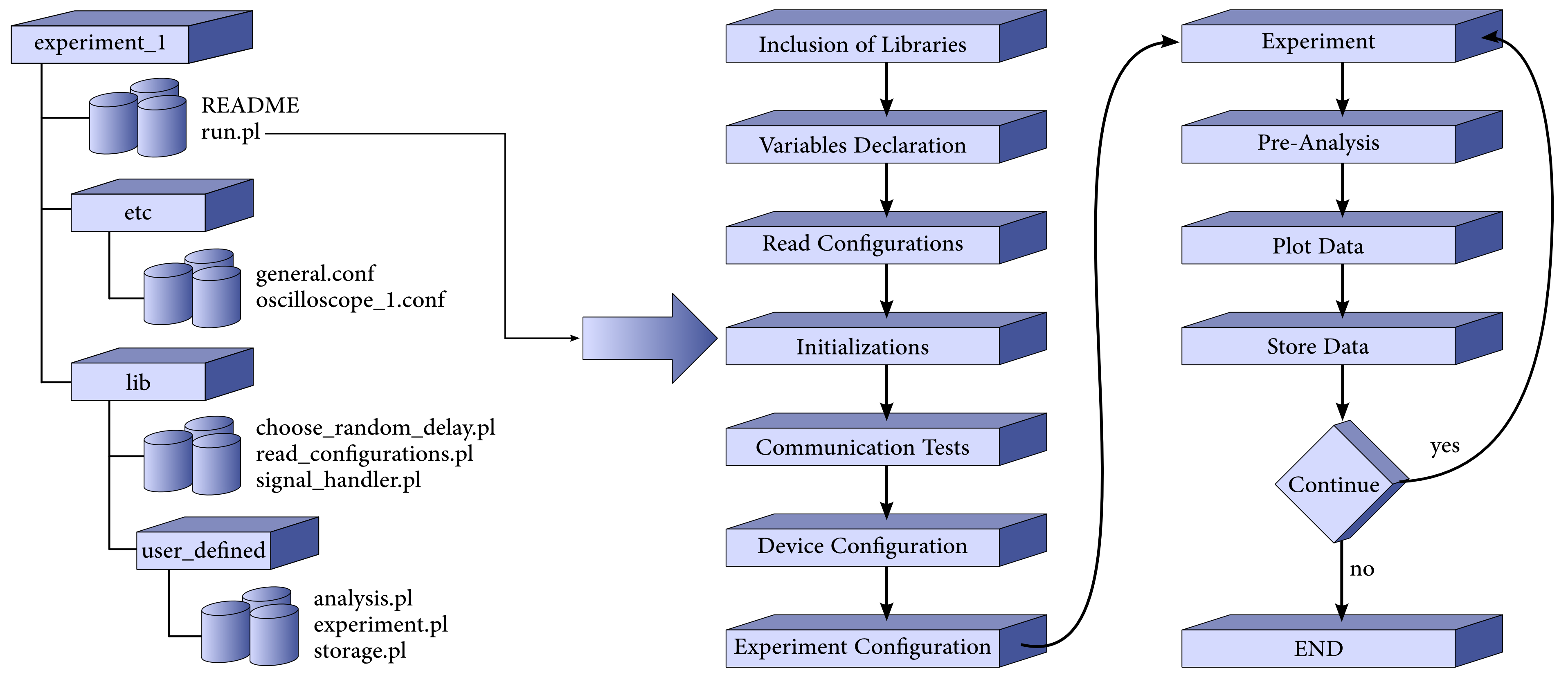}
  \caption{SCTE directory structure and experiment block diagram.}
  \label{fig:experiments}
\end{figure}

The directory tree of every example data acquisition system provided with SCTE is organized following the same scheme, shown on the left side of Figure~\ref{fig:experiments}. The root directory of the each experiment receives an unique name, e.g. {\tt experiment\_1}, {\tt experiment\_2}, or whatever name appropriated. These directories contain an experiment file ({\tt run.pl}), in which the data acquisition system is developed, a {\tt README} file with detailed information about the experiment, and at least two sub-directories, one for keeping configuration files ({\tt etc}), and one for libraries ({\tt lib}). 

The experiment configurations were split into two files, the {\tt general.conf}, that contains the most general configuration parameters required in the {\tt run.pl} file, and instrument specific configuration file (or files if more than one instrument is used) like {\tt oscilloscope\_1.conf}. All configuration files follow the same syntax i.e., the configuration parameters are pairs of keywords with their respective values separated by an equal sign ({\tt keyword=value}). The keywords may be defined according to specific demands of each data acquisition system. All these parameters are going to be read into a hash in the {\tt run.pl} application, facilitating its access and manipulation.

Using {\tt experiment\_1} as an example, some of the keywords defined in the {\tt general.conf} file are related to the instrument that is going to be manipulated in this specific application i.e., an USB oscilloscope. In this file (refer to the {\tt experiments/experiment\_1/etc/general.conf} file in the SCTE framework package for details), there are four keywords prefixed by {\tt DEVICE\_1} that relate to this instrument. These configuration parameters are used to inform the instrument BUS, path to the device file, delay between writes and reads, and the name of the configuration file specific to this instrument ({\tt oscilloscope\_1.conf}, in this case). If more than one instrument is being used in the specific application, new keywords may be added in order to provide the necessary configuration parameters for each device. This situation is presented in {\tt experiment\_2}, on which two instruments are employed, one USB and one RS-232.

Each run of the data acquisition system may have unique parameters associated, including the device being tested, the specific input signal frequency being applied, or even which output of the device is being tested. All this experiment specific information must be logged for later verification. In this way, the {\tt general.conf} file may contain several other experiment specific keywords in order to account for specific needs. As an example, the {\tt DEV\_ID\_UNDER\_CAL} keyword must receive a device identifier string that uniquely identifies the device being tested. It is possible to define as many keyword parameters as necessary, and they must be updated in each experiment.

The instrument configuration file syntax is similar to the general configuration file, except that the keywords must be SCPI command strings, and the values must be the values to be set in the instrument. As an example, the pair {\tt ACQ:MOD=SAM} will be understood as {\tt ACQ:MOD SAM}, which sets the acquisition mode of the oscilloscope as sampling. Since instruments may differ in purpose and functionality, although they use SCPI language, they may recognize different sets of commands thus requiring independent configuration files. Therefore, the programming manual of each specific instrument must be consulted. The experiments provided with SCTE were build for Tektronix TDS-1000 series oscilloscopes, whose SCPI commands, for example, differ from the Agilent DSO1012A oscilloscope, on which SCTE was also tested but required that most of the SCPI commands to be altered.

In the library directory there are two pre-defined libraries that are generic and applicable to any data acquisition system, and three additional user-defined libraries. The generic libraries defines two functions (one per file) the {\tt ReadConfigurations()} function, which reads configuration files and generates the appropriate hashes, and the {\tt SignalHandler()} function that simply handles process signals, such as {\tt SIGKILL}. The user defined libraries deal with the three major parts of data acquisition systems, the experiment instructions, pre-analysis of acquired data, and storage.

The proposed structure of a data acquisition system using SCTE is shown on the right side of Figure~\ref{fig:experiments}, which is implemented in the {\tt run.pl} application. As any other well structured computer program, the data acquisition system starts with the inclusion of external libraries, followed by variable definitions. The next step is to read the configuration files that provide the necessary information to continue with the execution. With the information contained in these files, it is then possible to initialize the interfaces with the instruments, initialize storage connections (connections with databases, or even simple file handlers initializations), and even auxiliary variables. With the instrument interfaces set up, we proceed with communication tests with the instruments in order to verify communication capability, allowing the identification of connection or configuration problems. Once stablished that all devices are reachable, the next step is to configure them to the purpose of the experiment by sending the appropriate SCPI commands contained in the instrument configuration file. The experiment configuration consists of using experiment specific information in order to prepare the {\tt run.pl} application to start the experiment. In the case of {\tt experiment\_1} the only required action is to initialize the vectors that will receive the data from the instrument.

After the preparation phase of the experiment, follows a loop that will perform the actual tasks relevant to the problem i.e., the control of the instruments by sending SCPI commands and reading instrument's  responses, thereof storing them into vectors. After each data acquisition cycle, the pre-analysis sub-routines perform appropriate calculations and display the results. If the results are acceptable i.e., there were no communication errors or any other event that invalidate the readings, the acquired data is stored for future more in-depth analysis. This loop continues until some stop condition is reached, or the user stops the experiment. Considering that these steps are the most complex ones in the development of this system, they were built as separated functions and split in three different library files under the {\tt lib/user\_defined} directory of the example experiment.

\section{Run test description}
\label{sec:run_tests}

\begin{figure}[H]
  \centering
    \includegraphics[width=0.6\textwidth]{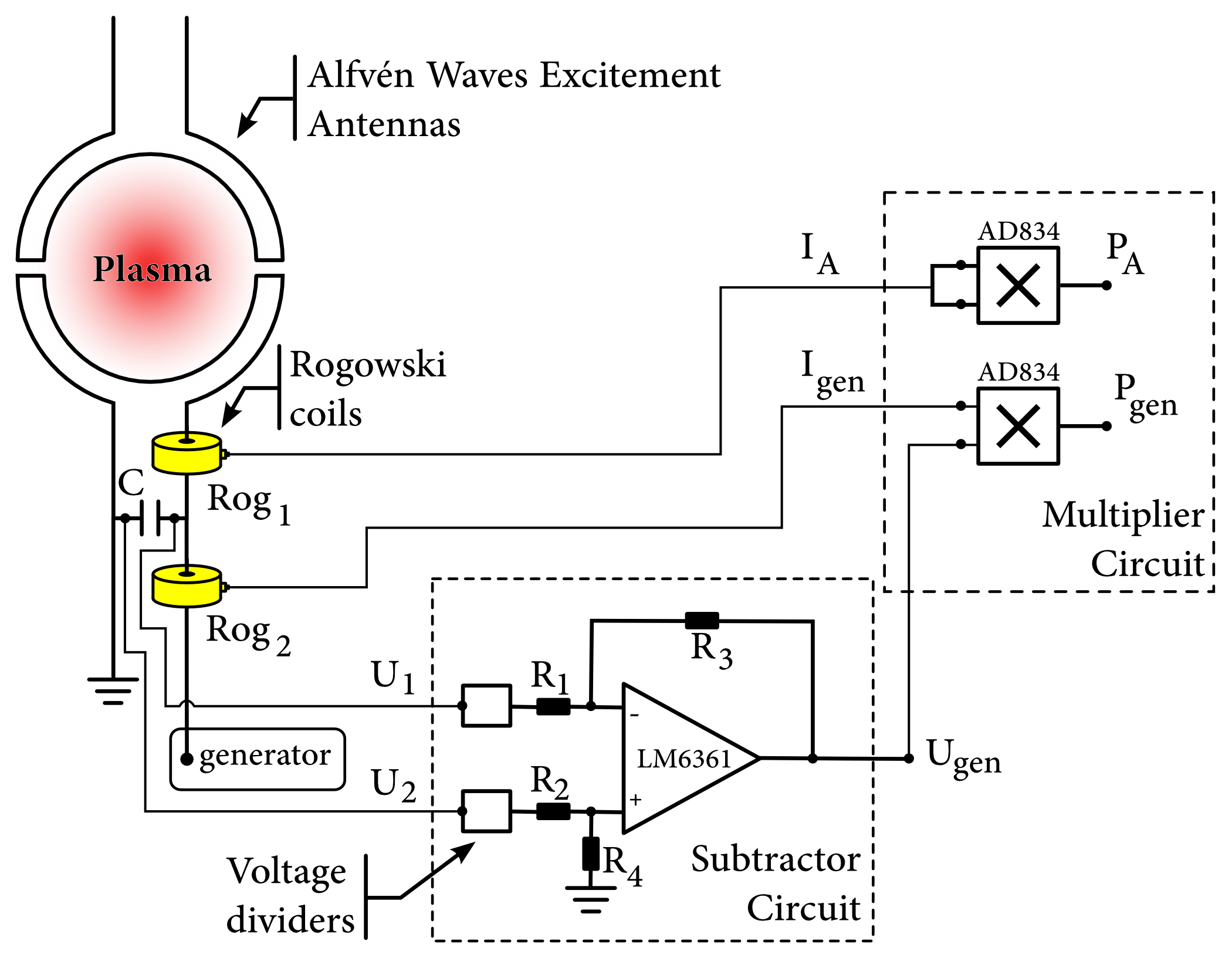}
  \caption{Power measurement infrastructure developed for the tokamak TCABR Alfvén Waves Excitement System.}
  \label{fig:awes}
\end{figure}

SCTE has been widely employed in the calibration of all in-house developed devices that integrates the power diagnostics of the AWES system. One set of antennas of the AWES system is capable of generating pulses of 100 kW of power at frequencies up to 6 MHz. Power measurements are performed by measuring the input current applied to the antennas by the generator using a Rogowski coil (Rog2, in Figure~\ref{fig:awes}), and at the same time, measuring the electric potential difference using a subtractor circuit based on an LM6361 operational amplifier, along with voltage divider for reducing the input voltage from 10 kV to approximately 1 V. Applying the radio-frequency (RF) current $I_{\mbox{gen}}$, and the RF voltage $U_{\mbox{gen}}$ signals to a multiplier circuit, the power delivered by the generator $P_{\mbox{gen}}$ can be determined.
By measuring the current $I_{A}$ in the antenna loop (via Rog1 coil), and multiplying it by itself using the multiplier circuit (based on an AD834 four-quadrant multiplier), the power losses in the antenna loop $P_A$ can be determined. Therefore, the difference between $P_{\mbox{gen}}$ and $P_A$ determines the power dissipated in the plasma, after proper calibrations.

\begin{figure}[H]
  \centering
    \includegraphics[width=0.7\textwidth]{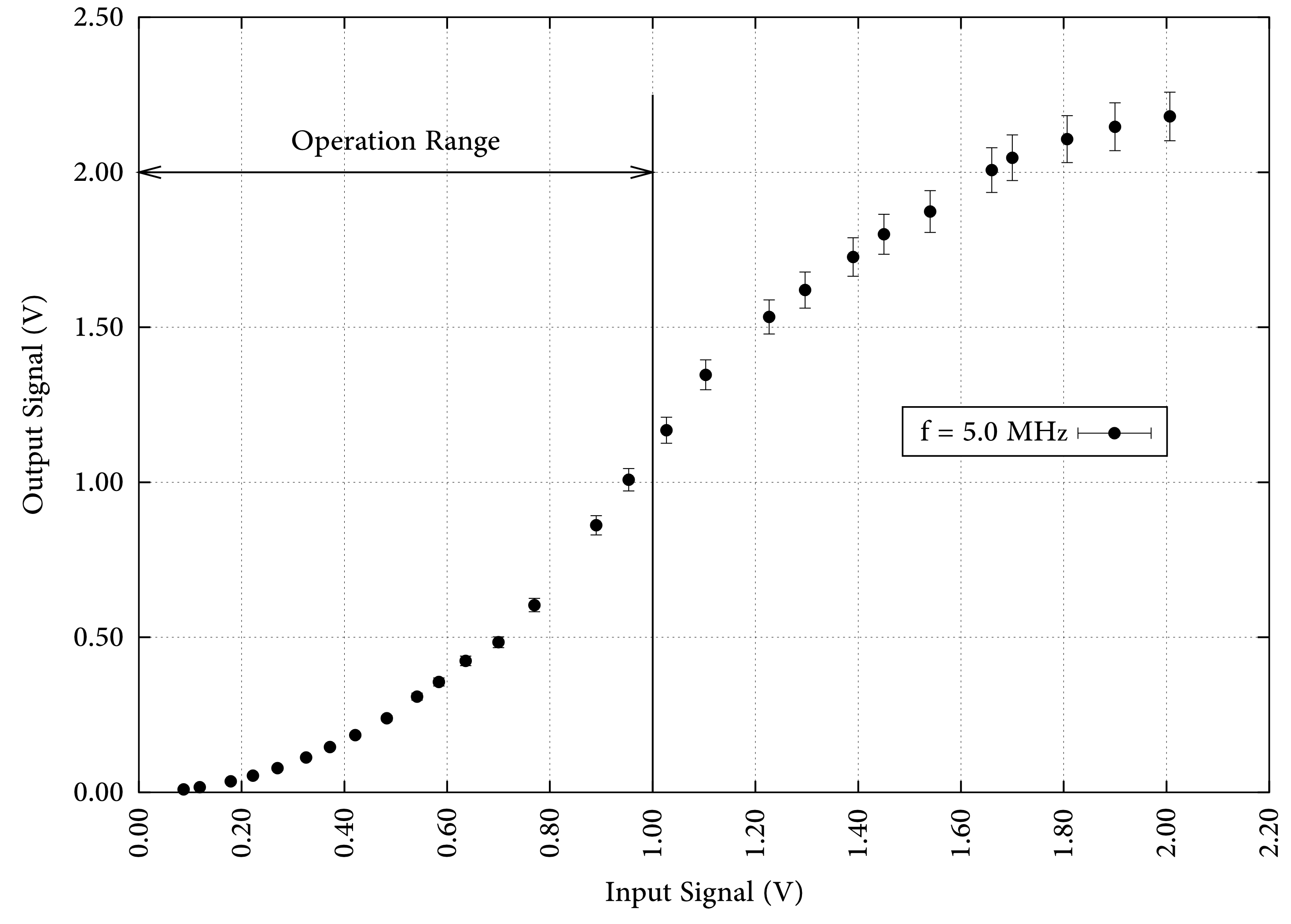}
  \caption{Calibration curve for the RF signal multiplier resulting from an SCTE automated experiment. Each point correspond to {\tt READINGS\_PER\_POINT} readings per state, and the error bars corresponds to the standard error.}
  \label{fig:archives_figs_paper1}
\end{figure}

The first and second experiments (namely {\tt experiment\_1} and {\tt experiment\_2}) are examples of data acquisition systems that perform the calibration of devices. They were used to determine the calibration curve of the Rogowski coils, and of the subtractor and multiplier circuits, as an example. Figure~\ref{fig:archives_figs_paper1} shows the 
calibration curve of one of the multiplier circuits. In order to produce this curve, an input signal of approximately 5 MHz was applied to both inputs of the multiplier, varying the amplitude of the signal from 0 to 2 V. Since the same input signal was applied to both input ports of the multiplier, the resulting signal should be the simple multiplication of the input signal by itself, which is valid within the operation range of the AD834 component, which is true for signals up to 1 V.

The input signal was generated using an HP3112A generator, which had no control ports. The measurements were performed with a Tektronix TDS-210 oscilloscope, in which channel one was used to measure the input signal, and channel two was used to measure the output signal. The oscilloscope was controlled by SCTE acquiring the peak to peak value and signal frequency on both channels. In each step, the amplitude of the input signal was set and the measurements performed in a sequence of events defined in the {\tt RunExperiment()} function. Initially, the oscilloscope was set into stop mode and the first channel was selected performing frequency and peak to peak measurements. In the sequence, channel two was selected and the same measurements were performed. After the completion of the first round of measurements, the oscilloscope was set to run for a small interval between 0 and 1 second, after which the measurement procedure was repeated, thus generating a new set of readings for the same input signal. This procedure was repeated by {\tt READINGS\_PER\_POINT} times, as specified in the {\tt general.conf} file, in order to account for possible variations. 

When the specified number of repeated readings was reached, the {\tt AnalyseReadings()} sub-routing was called, and average values, standard deviations, and standard errors were calculated and graphically presented to the user. After the evaluation of the results by the end-user, a new amplitude of the input signal was set, and SCTE was instructed to continue. With the acceptance of the previous acquired data, SCTE executed the {\tt WriteData()} function in order to store the recently acquired data, and then executed the {\tt RunExperiment()} again, initiating a new cycle of readings.

\begin{figure}[H]
  \centering
    \includegraphics[width=1\textwidth]{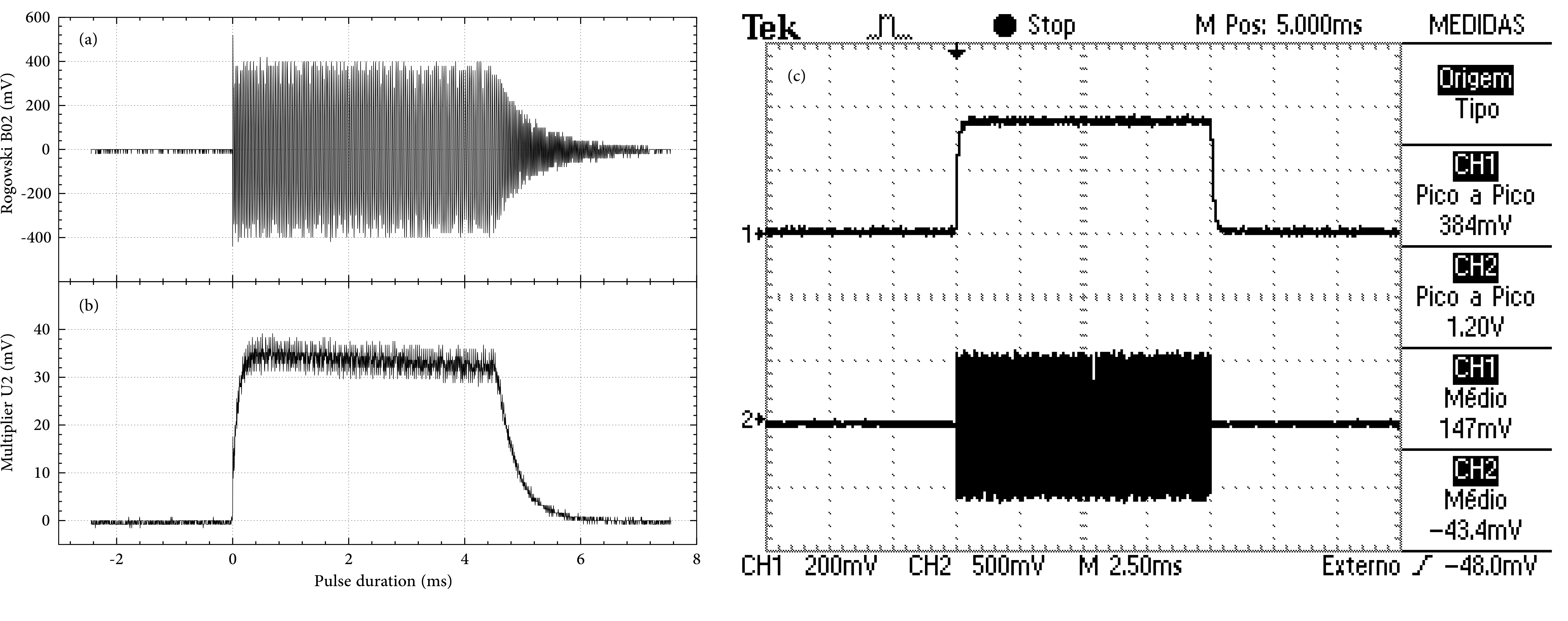}
  \caption{Example of outputs from experiments 3 and 4. (a) and (b) correspond to wave forms captured with {\tt experiment\_3}, and (c) corresponds to a screen shot of the oscilloscope screen captured with {\tt experiment\_4}.}
  \label{fig:archives_figs_two_signal}
\end{figure}

Additionally to experiments 1 and 2, which perform the same task but using one and two instruments respectively, experiments 3 and 4 present different applications of SCTE. Experiment 3 ({\tt experiment\_3}), presents a simple application to acquire the wave form data points of the signal displayed in the oscilloscope screen (Figures~\ref{fig:archives_figs_two_signal} a and b), while experiment 4 ({\tt experiment\_4}) implements the snapshot functionality that captures an image of the oscilloscope screen (Figure~\ref{fig:archives_figs_two_signal}c).

Two wave forms originated from one TCABR test shot are shown in Figures~\ref{fig:archives_figs_two_signal}a and b. The first acquired from a Rogowski coil (B02) and used as input to the multiplier system, and second, the resulting multiplication. Using the sensitivity factor of the Rogoswki coil B02, which is 18.61(16)$\times 10^{-3} \frac{V}{A}$, we conclude that the measured current was approximately 20 A during the pulse. Using the proportionality constant for the multiplier U2, which is approximately 121 $\frac{W}{mV}$, we conclude that the power was kept at approximately 4.1 kW during the pulse.

In order to keep a reference of the operation behaviour of one the multipliers, a screenshot was acquired during its calibration process, which is presented in Figure~\ref{fig:archives_figs_two_signal}c. In order to simulate the RF current and voltage signals, two amplitude modulated signals, with carrier frequency of approximately 5 MHz and duration of 10 ms were supplied as input to the multiplier circuit and to the oscilloscope as reference signal. The reference signal was measured in channel 2, and the output of the multiplier was captured in channel 1. This screenshot can be used as a reference of the operation behaviour of the referred equipment for future consultation. These four examples cover the basic usage of the SCTE framework, and provide examples on how to use it while employing general concepts of the development of data acquisition systems.

\section{Conclusions}  

SCTE constitutes a simple, yet powerful, framework for creating automated data
acquisition and equipment control systems for experimental Physics and related areas. The {\tt SCTE::Instrument} module provides an effective layer of communication with instruments via RS-232, USB, and Ethernet (LAN). Although SCTE experiment applications cover the most common operations with oscilloscopes, SCTE is not limited to oscilloscopes and can be used to control any SCPI enabled instrument, which makes it a very versatile and cost-effective option.

\section{Acknowledgements}  

The authors would like to thank the reviewers for their insightful comments that
have improved this paper. This work received financial support 
from the Brazilian agency CNPq.





\bibliographystyle{elsarticle-num}
\bibliography{references}

\end{document}